\documentclass[runningheads]{llncs}

\usepackage{eccv}

\usepackage{eccvabbrv}

\usepackage{graphicx}
\usepackage{booktabs}
\usepackage{multirow}
\usepackage{amssymb}
\usepackage{colortbl}
\usepackage{tikz}
\usepackage{ctable}
\usepackage{pifont}
\newcommand{\cmark}{\ding{51}}%
\newcommand{\xmark}{\ding{53}}%

\usepackage[accsupp]{axessibility}  

\usepackage{hyperref}

\usepackage{orcidlink}

\begin{document}

\title{Quanta Video Restoration} 

\titlerunning{QUIVER}

\author{Prateek Chennuri\inst{1} \and
Yiheng Chi\inst{1} \and
Enze Jiang\inst{1} \and G.M. Dilshan Godaliyadda\inst{2} \and Abhiram Gnanasambandam\inst{2} \and Hamid R. Sheikh\inst{2} \and \\ Istvan Gyongy\inst{3} \and Stanley H. Chan\inst{1}}

\authorrunning{P.~Chennuri et al.}

\institute{Purdue University \and
Samsung Research America \and University of Edinburgh\\
\email{\{pchennur,chi14,jiang708,stanchan\}@purdue.edu}\\
\email{\{dilshan.g,abhiram.g,hr.sheikh\}@samsung.com}\\
\email{\{igyongy2\}@exseed.ed.ac.uk}
}

\maketitle
\begin{center}
  \centering
  \captionsetup{type=figure}
   \includegraphics[width=0.95\linewidth]{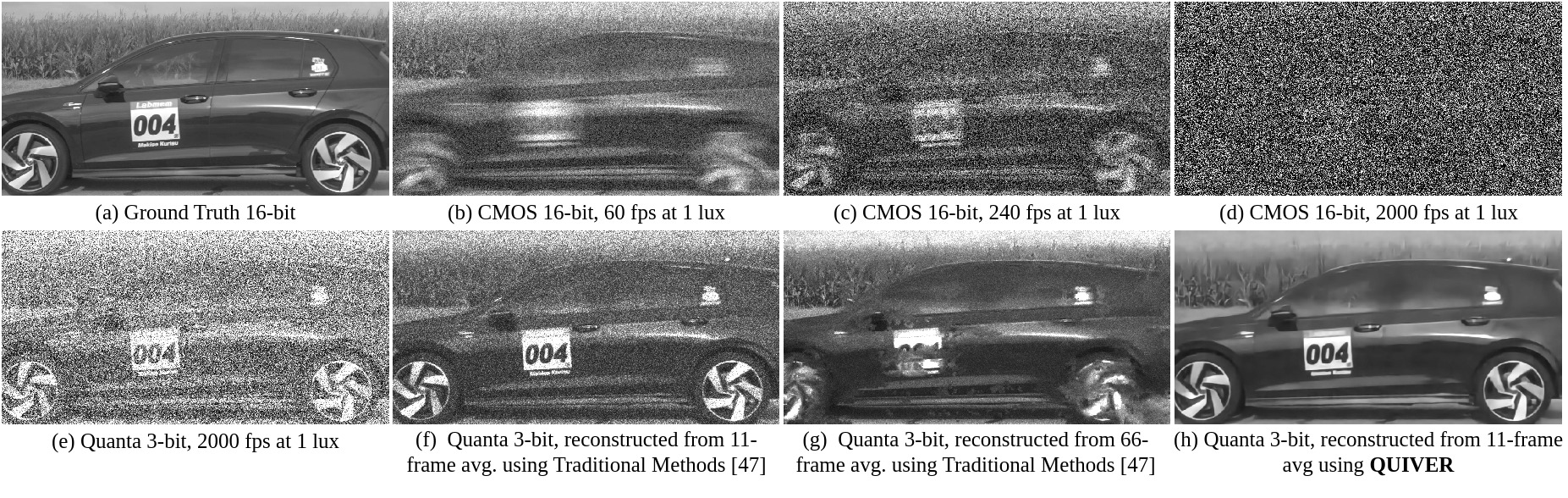}
   \captionof{figure}{\textbf{Goal of this paper}. (a) Blur-free video frame of a moving car. (b)-(d) CMOS image sensor simulations using realistic sensor parameters. The strong shot noise and read noise ($5.1\;\text{e}^-$/pix) of CMOS sensor make the signal acquisition difficult. (e) With low read noise ($0.2\;\text{e}^-$/pix), low-bit single-photon detectors capture valuable information. (f)-(g) Existing state-of-the-art algorithm, QBP~\cite{maQuantaBurstPhotography2020} cannot handle strong motion and noise. (h) The proposed algorithm, QUIVER, produces high quality results.} 
   \label{fig:paper_fig1}
\end{center}
\begin{abstract}
  The proliferation of single-photon image sensors has opened the door to a plethora of high-speed and low-light imaging applications. However, data collected by these sensors are often $1$-bit or few-bit, and corrupted by noise and strong motion. Conventional video restoration methods are not designed to handle this situation, while specialized quanta burst algorithms have limited performance when the number of input frames is low. In this paper, we introduce Quanta Video Restoration (QUIVER), an end-to-end trainable network built on the core ideas of classical quanta restoration methods, i.e., pre-filtering, flow estimation, fusion, and refinement. We also collect and publish I$2$-$2000$FPS, a high-speed video dataset with the highest temporal resolution of $2000$ frames-per-second, for training and testing. On simulated and real data, QUIVER outperforms existing quanta restoration methods by a significant margin. Code and dataset available at \url{https://github.com/chennuriprateek/Quanta_Video_Restoration-QUIVER-}
\end{abstract}

\section{Introduction}
\label{sec:intro}
Over the past decade, the astonishing growth of single-photon detectors has fundamentally changed the landscape of computational imaging. With the invention and proliferation of quanta image sensors (QIS)~\cite{fossumQuantaImageSensor2016} and  single-photon avalanche diodes (SPAD)~\cite{rochasalexisSinglePhotonAvalanche2003, niclassDesignCharacterizationCMOS2005},
there is an unprecedented volume of new applications in low-light imaging~\cite{chiDynamicLowLightImaging2020, chanImagesBitsNonIterative2016b, shinPhotonefficientImagingSinglephoton2016}, computer vision~\cite{gnanasambandamImageClassificationDark2020, liPhotonlimitedObjectDetection2021, guptaEulerianSinglePhotonVision2023}, high-speed videography~\cite{maQuantaBurstPhotography2020, maBurstVisionUsing2023}, time-of-flight sensing~\cite{gutierrez-barraganCompressiveSinglePhoton3D2022, rugetPixels2PoseSuperresolutionTimeofflight2022}, and 3D imaging~\cite{guptaPhotonFloodedSinglePhoton3D2019, lindellSinglephoton3DImaging2018}. In most of these use cases, the main core question that lies is how to recover the image from the photon counts measured in the scene. Specifically, given a video stream of $1$-bit or few-bit data captured from a scene involving moving objects, how do we reconstruct a gray-scale image/video while eliminating the noise without incurring motion blur?

To give the reader a visual perspective of the problem scope, \cref{fig:paper_fig1} depicts a blur-free video of a moving car.  We simulate the captured images at 1 lux assuming $60$ fps, $240$ fps, and $2000$ fps CMOS image sensors with realistic sensor specifications. As illustrated in the figure, the resulting CMOS outputs are either severely blurred due to strong motion or completely distorted by noise due to sparse photons. In the same figure, we demonstrate a simulated single-photon camera output (a $3$-bit QIS in this case) where the content the largely preserved despite heavy noise. Upon utilizing state-of-the-art Quanta Burst Photography (QBP)~\cite{maQuantaBurstPhotography2020} for reconstructing the frames, provided the motion is slow, a decent output can be obtained. However, as the temporal window narrows down, as shown in \cref{fig:paper_fig1} (f), the noise remains. In this paper, we address this problem with a new algorithm, designed to remove the noise while avoiding distortions in the presence of fast motion while utilizing only a few frames.

The core innovation of this paper is QUanta VIdeo REstoration (QUIVER), a deep-learning based video restoration algorithm for quanta image data. QUIVER is specialized for few-bit data ($3$-bit) captured at thousands of frames-per-second ($2000$ fps) with an average motion range of $1$ to $7$ pixels per frame. The main contributions of this paper can be summarized as follows.
\begin{itemize}
    \item We propose QUIVER, an end-to-end trainable quanta (video) restoration method built by embracing the core ideas from traditional quanta restoration algorithms. On a comprehensive evaluation dataset containing both simulated and real data, QUIVER outperforms all methods we compared in this paper by a significant margin. 
    \item We introduce I$2$-$2000$FPS, the first high-speed video dataset with a temporal resolution of $2000$ frames-per-second for training and testing image and video reconstruction neural networks. We captured a total of $280$ high-speed videos covering $114$ distinct scenes with ground truth and simulated $3$-bit videos. 
\end{itemize}

\section{Background}
\label{sec:background}
\begin{figure}[tb]
  \centering
   \includegraphics[width=0.8\linewidth]{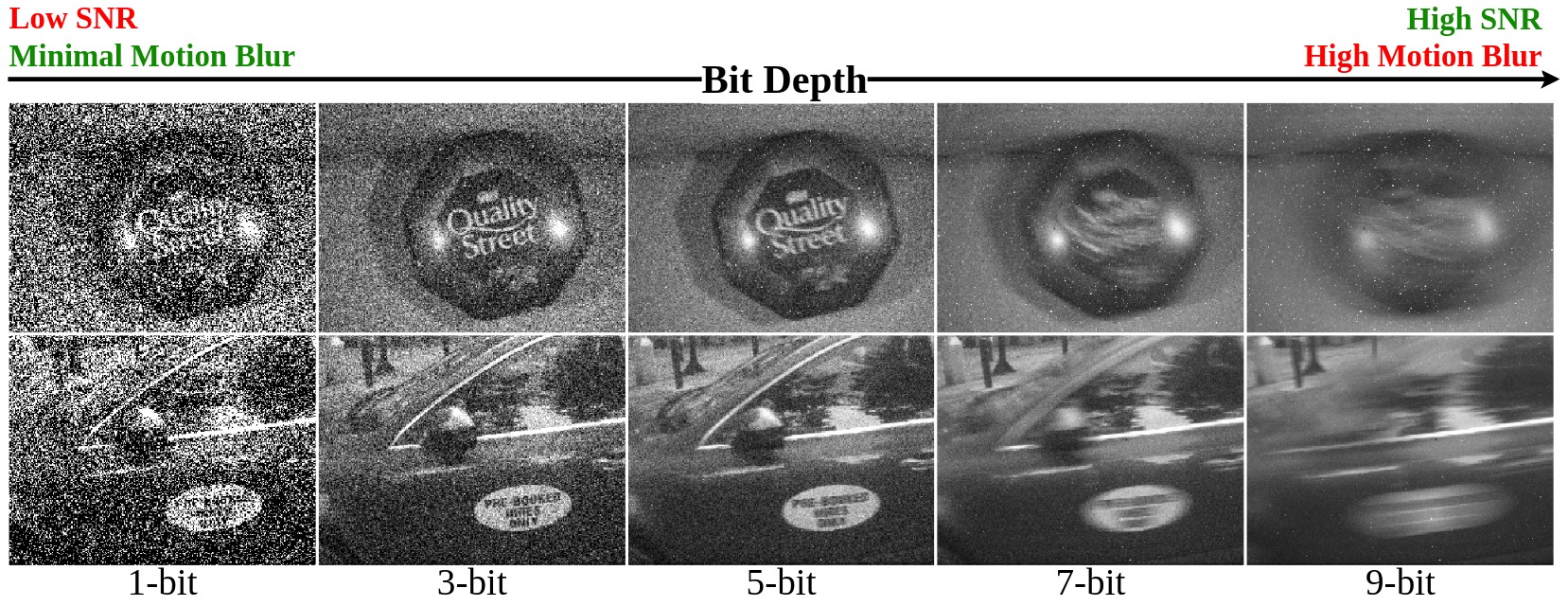}
   \vspace{-1ex}
   \caption{\textbf{Motion Blur and SNR Trade-off}. The effects of bit depth on SNR and motion blur are illustrated using \emph{real} captures by a single-photon sensor. For the motion range we target, $3$-bit single-photon detectors provide the best trade-off between blur and SNR. The images are captured using a $1$-bit SPAD~\cite{duttonSPADBasedQVGAImage2016} at $10$k fps at an average photon level of $0.51$ and $0.40$ photons-per-pixel (PPP) per frame, respectively. Higher bit-depth outputs are generated through temporal frame averaging.
   }
   \label{fig:bitdepth_SNR_motionblur}
\end{figure}
\begin{table}
\caption{Frame-rate, motion, read-noise, and data-rate statistics for various bit-depths at the same exposure level.}
    \centering
    \setlength{\arrayrulewidth}{.13em}
    \resizebox{0.5\linewidth}{!}{
    \begin{tabular}{ccccc}
    \toprule[2pt]
        \multirow{2}{*}{Bit-Depth} & \multirow{2}{*}{fps} & Motion & $\sigma_{\text{read}}$ & Data-rate\\
         &  & (pixels/frame) & (/pixel/sec) & (Mb/sec)\\ 
         \midrule[1pt]
        $1$ & $10$k & $0-1$ & $2000\;\text{e}^-$ & $96$ \\
        $3$ & $1428$ & $2-3$ & $285.6\;\text{e}^-$ & $41.13$ \\
        $5$ & $323$ & $6-12$ & $64.6\;\text{e}^-$ & $15.5$\\
        $7$ & $78$ & $25-30$ & $15.6\;\text{e}^-$ & $5.24$\\
        $9$ & $20$ & $70-80$ & $4\;\text{e}^-$ & $1.73$\\
        \bottomrule[2pt]
    \end{tabular}
    }
    \label{tab:bit_depth_comp}
\end{table}
\subsection{Few-bit Single-Photon Detectors}
\textbf{What is few-bit photon counting?} Single-photon detectors (QIS and SPAD) differ from conventional CMOS pixels by their extraordinary photon counting capability. QIS uses a two-stage pump-gate technique and correlated double sampling to suppress the read noise, while SPAD uses avalanche multiplication to amplify the photocharge. In both cases, the sensors are capable of resolving photons up to a single-photon sensitivity. We refer readers interested in the sensor development of QIS and SPAD to consult, for example,~\cite{chanWhatDoesOneBit2022, maReviewQuantaImage2022a, charbonSPADBasedSensors2013, duttonSinglePhotonCounting2016}.

Along with the single-photon detectors' unique capability to count individual photons, these devices can generate data at a bit-depth as low as $1$-bit to as high as $16$-bit or even more. However, higher bit-depth is accompanied by longer integration time. If the scene contains motion, longer integration time will eventually result in strong motion blurs as shown in \cref{fig:bitdepth_SNR_motionblur}. On the other hand, $1$-bit sensing with high frame rates will result in motion-blur-free but extremely noisy images. Therefore, from a pure data acquisition perspective, there exists an optimal bit-depth with respect to the motion that will give us minimal/no motion-blur data with a minimum per-frame signal-to-noise ratio (SNR) required for good quality reconstruction.

\textbf{How about 1-bit and reconstruct afterward?} Readers familiar with single-photon counting may wonder whether we can collect as many $1$-bit frames as possible and then process the data afterward. The problem is power consumption and data rate. Fixing the same level of exposure, as described in \cref{tab:bit_depth_comp}, a $1$-bit video at $10$k fps would require $96$ Mb/sec whereas a $9$-bit video at $20$ fps would only need $1.73$ Mb/sec. Another problem is read noise accumulation. For sensors with non-zero read noise (such as QIS), every frame contributes to a finite amount of read noise. The more frames we read, the more read noise we accumulate. Therefore, recording $1$-bit is not always the best option.

\subsection{Related Work}
\label{subsec:related_work}
\textbf{Image and Video Denoising}. Classical state-of-the-art methods utilize a non-local strategy to identify similar patches across an image/video~\cite{maggioniVideoDenoisingDeblocking2012, lebrunNonlocalBayesianImage2013, ariasVideoDenoisingEmpirical2018}. Deep neural networks have been proven to be successful in producing high quality denoised outputs~\cite{clausViDeNNDeepBlind2019, tassanoDVDNETFastNetwork2019,tassanoFastDVDnetRealTimeDeep2020, vaksmanPatchCraftVideo2021}. Among these architectures, Vision Transformers~\cite{liangVRTVideoRestoration2022, liangRecurrentVideoRestoration2022a} have been rated the state-of-the-art in recent times. However, all these solutions make simplistic assumptions on noise statistics, thus failing to perform on real noisy images or videos~\cite{plotzBenchmarkingDenoisingAlgorithms2017}.

Coming to low light, Burst denoising~\cite{hasinoffBurstPhotographyHigh2016b, libaHandheldMobilePhotography2019}, where images are aligned, merged and denoised, is one of the most popular methods. However, these methods fail without robust alignment. To overcome this, a number of alternative solutions with learnable alignment modules have been proposed~\cite{godardDeepBurstDenoising2018, vogelsDenoisingKernelPrediction2018,pearlNANNoiseAwareNeRFs2022}. Recent solutions have focused on practical noise models that replicate real camera sensor noise, to produce visually appealing results~\cite{monakhovaDancingStarsVideo2022, wangEnhancingLowLight2019}. Nevertheless, the  existing solutions utilize images captured using CMOS image sensors, resulting in a notably higher photon level compared to the one utilized in our study. 

\textbf{SPADs, Event, and Spike Cameras.} 
Gariepy \etal \cite{gariepySinglephotonSensitiveLightinfight2015, gariepyPicosecondTimeresolvedImaging2016} firstly introduced the utilization of Single Photon Avalanche Diodes (SPADs) at pico-second temporal resolution to capture light in motion. Gyongy \etal~\cite{gyongySinglePhotonTrackingHighSpeed2018} demonstrated $2$D motion tracking of rigid planar objects using SPADs at $10$k fps. Recently, Ma \etal~\cite{maQuantaBurstPhotography2020} and Seets \etal~\cite{seetsMotionAdaptiveDeblurring2021} utilized SPADs in a passive imaging setting to capture motion in low illumination. However, all these methods utilize extremely high-temporal resolutions hindering the deployment of these sensors into consumer devices where bandwidth is the bottleneck. Event~\cite{panBringingBlurryFrame2019, rebecqHighSpeedHigh2021} and Spike Cameras~\cite{zhaoHighSpeedMotionScene2020, zhaoReconstructingClearImage2022} also have demonstrated their effectiveness in capturing high-speed motion. However, these cameras focus on luminance/brightness variation and record a spike only when variation is above a threshold (changes based on factors like temperature, event rate, etc.)~\cite{dong2021spike, rebecqHighSpeedHigh2021}. Therefore, unlike single-photon detectors (QIS and SPADs), these cameras are NOT designed for single-photon counting and cannot operate in extreme low-light conditions.

\textbf{QIS Reconstruction.} Reconstructing quanta images is a challenging task due to the underlying Poisson-Gaussian statistics. Initial solutions to this problem included utilizing standard gradient descent~\cite{yangOptimalAlgorithmReconstructing2010}, greedy algorithms~\cite{yangImageReconstructionGigavision2009}, (ADMM)~\cite{chanEfficientImageReconstruction2014, chanPlugandPlayADMMImage2017}, among others~\cite{fengyangBitsPhotonsOversampled2012, remezPictureWorthBillion2016, elgendyOptimalThresholdDesign2018, gnanasambandamHDRImagingQuanta2020, wongTheoreticalAnalysisImage2021, gaoHighDynamicRange2022}. Chan \etal~\cite{chanImagesBitsNonIterative2016b} were the first to propose a non-iterative approach using Anscombe transform for reconstructing quanta images. Choi \etal~\cite{choiImageReconstructionQuanta2018} proposed the first end-to-end trainable deep neural network (DNN) for QIS reconstruction. Alternative DNN-based solutions include utilizing vision transformers~\cite{wangSinglePhotonCamerasImage2023}, Dual Prior Integrated networks~\cite{zhangDualPriorIntegratedImage2023}, among others~\cite{zhangTwPTwostageProjection2023}. Nonetheless, all these methods fail to produce good results when the scene is in motion. Chi \etal~\cite{chiDynamicLowLightImaging2020} is the only method which focuses on capturing dynamic scenes using QIS but only targets at slow motion ($1$ pixel/frame).

\section{QUanta VIdeo REstoration (QUIVER)}
\label{sec:quiver}
\subsection{Design Philosophy}
\begin{figure}[tb]
  \centering
   \includegraphics[width=0.95\linewidth]{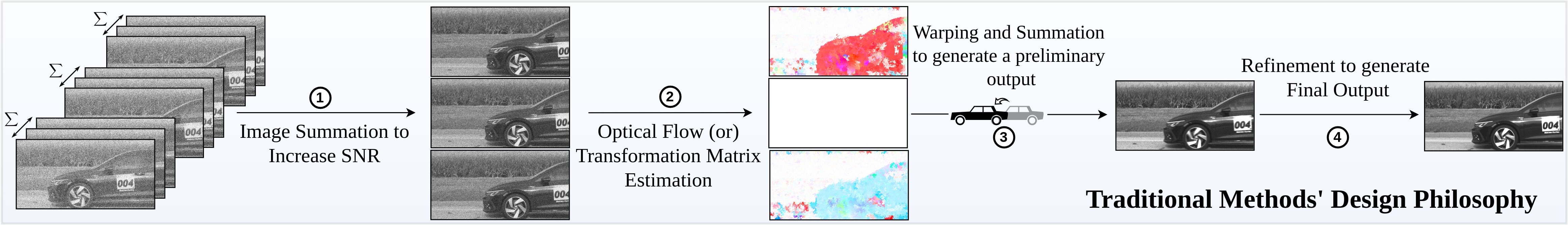}
   \caption{\textbf{Traditional Methods' Design}. Depiction of existing classical quanta restoration algorithms' design philosophy. \textit{Best viewed in zoom.}}
   \label{fig:trad_meth_design}
\end{figure}

In this section we present the design of our proposed algorithm.
We start by briefly reviewing the design of classical methods~\cite{gyongySinglePhotonTrackingHighSpeed2018, maQuantaBurstPhotography2020} which, to an extent, have been successful in restoring quanta images. As shown in \cref{fig:trad_meth_design}, classical methods' algorithm design can be divided into four stages: $(1)$ computing sum images to increase SNR, $(2)$ optical flow (or) transformation matrix estimation for aligning the input frames, $(3)$ warping and linear combination for generating preliminary restored output, and $(4)$ refinement for producing the final output. While the steps seem intuitive and straightforward, existing methods are heavily vulnerable to extreme noise and strong motion in the input frames primarily due to two reasons. ($1$) none of the stages are designed to handle extreme noise and strong motion simultaneously (will be discussed further). ($2$) Since all the stages are sequential yet independent of each other, it is difficult to obtain an optimal result for a wide range of noise and motion. Our proposed algorithm QUIVER, leverages the design philosophy of existing classical methods while designing each stage to simultaneously handle both noise and motion. Moreover, QUIVER is an end-to-end trainable model making all the stages inter-dependent, thus leading to good restoration outputs.  
\subsection{Design of QUIVER}
\begin{figure}[tb]
  \centering
   \includegraphics[width=\linewidth]{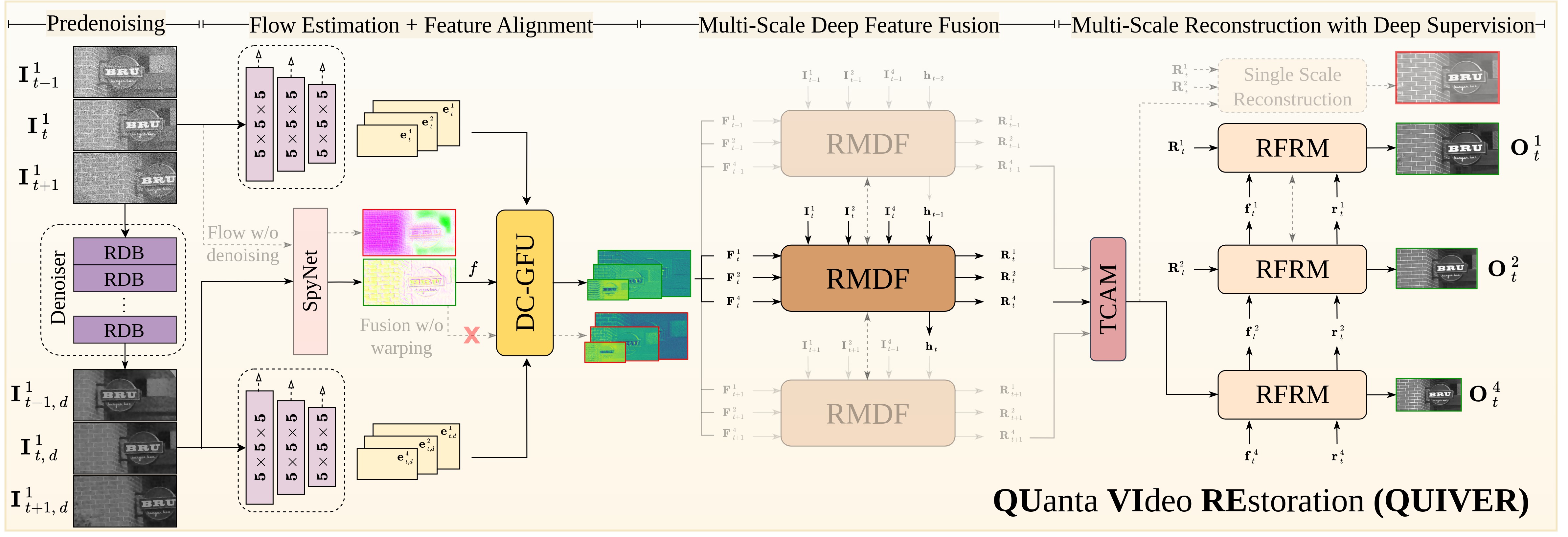}
   \caption{\textbf{The proposed QUIVER network}. The corresponding stages of QUIVER, built by embracing the intuitive thoughts behind existing classical methods. \textit{Best viewed in zoom.}}
   \label{fig:quiver_outline}
\end{figure}

QUIVER is a deep-learning-based video restoration method for quanta imaging. The design philosophy of QUIVER is to adopt the intuitive thoughts behind classical quanta restoration methods and develop an end-to-end trainable, robust to noise and motion deep learning based framework, as shown in \cref{fig:quiver_outline}. Specifically, QUIVER can be divided into four main stages:
\begin{figure}[tb]
  \centering
  \includegraphics[width=0.9\linewidth]{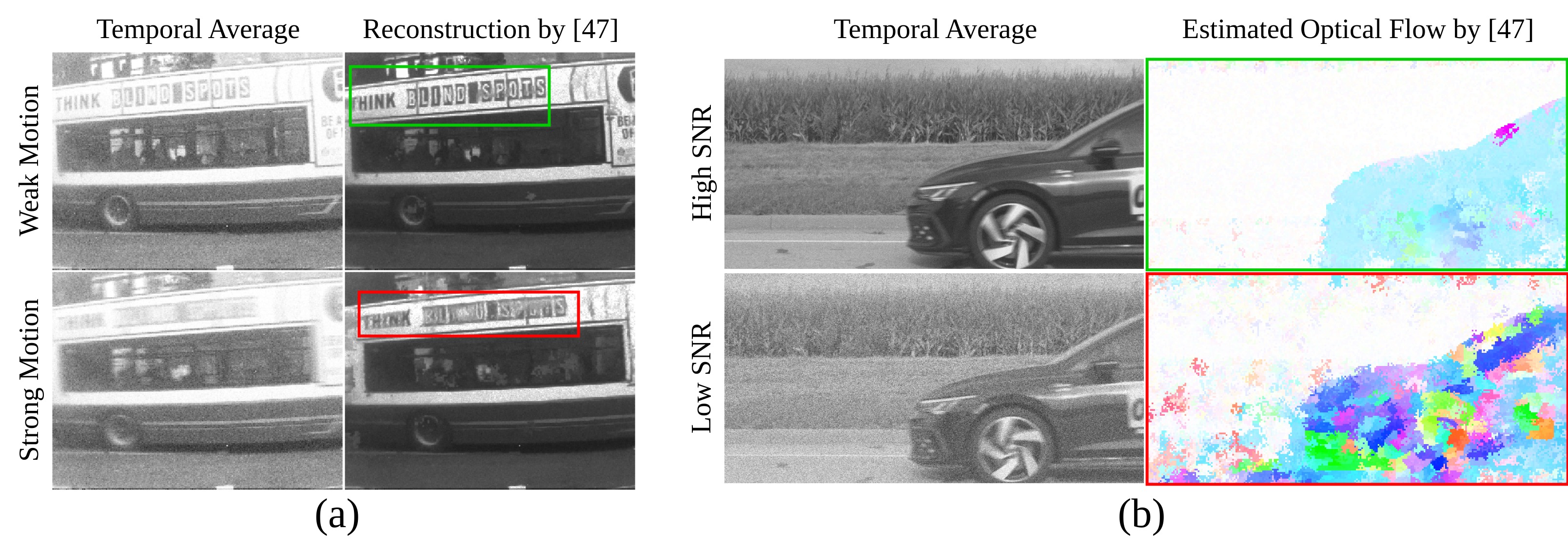}
   \caption{\textbf{Traditional Methods' Limitations}. (a) Traditional methods'~\cite{maQuantaBurstPhotography2020,gyongySinglePhotonTrackingHighSpeed2018} predenoising/temporal-averaging fails in strong motion. It is visible in the restored images that an input with strong motion between the frames results in several artifacts in the output even though SNR levels are similar. (b) Traditional methods~\cite{maQuantaBurstPhotography2020} utilize a patch-based pre-trained optical flow module similar to~\cite{hasinoffBurstPhotographyHigh2016b}. This optical flow module fails to compensate for motion in the presence of noise.}
   \label{fig:qbp_lim}
\end{figure}

\textbf{Pre-Denoising to improve SNR}: 
Since the input quanta frames possess extreme noise, classical methods adopt naive averaging to increase the SNR and thereby predict better optical flows or transformation matrices. 
However, as shown in \cref{fig:qbp_lim}(a), the simple averaging is vulnerable to motion and will negatively impact subsequent processing, ultimately leading to distorted outputs.  
Simply eliminating this stage is not the solution, 
because it leads to poor optical flow estimation, resulting in over-smoothed outputs with lack of low-level intricate details, as shown in \cref{fig:quiver_outline} and \cref{fig:stages_importance}. Therefore, a preliminary denoising step robust to noise and motion is crucial. 
In QUIVER, we leverage a computational undemanding single image denoiser built using Residual Dense Blocks~\cite{zhangResidualDenseNetwork2021} (RDBs) to provide minimal pre-preprocessing of the input quanta data. A multi-frame denoiser is not an option due to it computationally demanding nature. We use RDB based network due to its history in handling noise while preserving details with its simple yet effective design. 

\textbf{Optical Flow Estimation + Feature Alignment using Deformable Convolution - Gated Fusion Unit}: Classical methods utilize an off-the-shelf pre-trained optical flow estimation module or predict a transformation matrix to compensate for motion between the frames. The basic assumption behind such approaches is that the motion between the frames is limited and the SNR is high enough. However, when such assumption is not met, the motion compensation is sub-optimal, as shown in \cref{fig:qbp_lim}(b). As most state-of-the-art pre-trained optical flow estimators are optimized on the CMOS RGB sensor images, it leads to sub-optimal performance when applied on quanta frames. Eliminating the flow estimation step is not recommended since experiments reveal the critical role it plays in motion compensation, as shown in \cref{fig:quiver_outline} and \cref{fig:stages_importance}.
For QUIVER, we deploy a learnable optical flow estimation module and utilize SpyNet~\cite{ranjanOpticalFlowEstimation2017} owing to its computational efficiency while using a multi-scale approach.
    
We deploy $3$D convolution blocks to extract multi-scale spatio-temporal features from both the noisy and denoised quanta frames. We reuse the noisy frames to compensate for any information lost in the pre-denoising stage. 

\begin{figure}[tb]
  \centering
   \includegraphics[width=\linewidth]{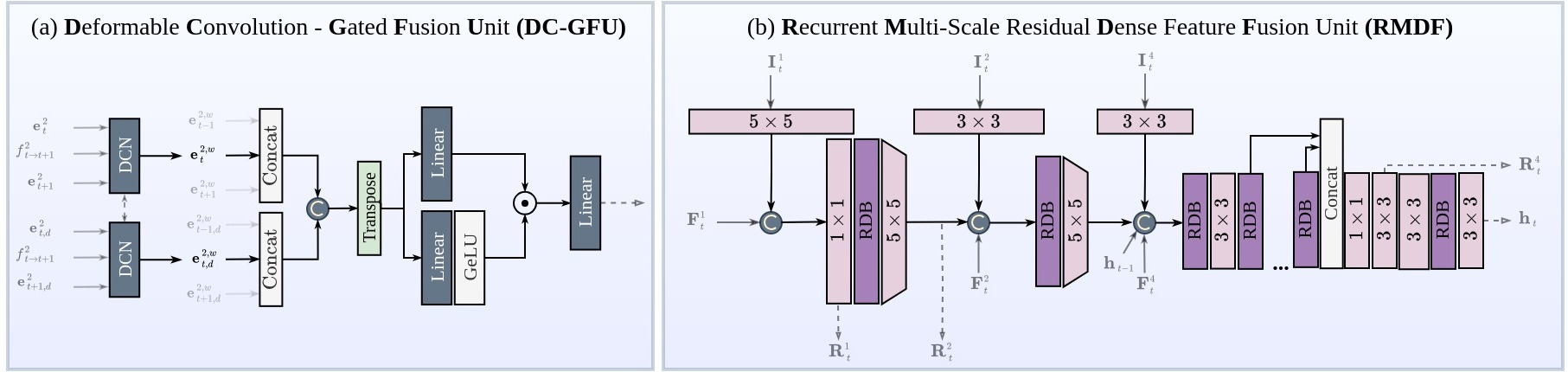}
   \caption{Design of the proposed modules DC-GFU and RMDF.}
   \label{fig:dcgfu_rmdf}
\end{figure}

The estimated multi-scale robust-to-noise optical flows are utilised for feature-level alignment of the extracted multi-scale spatio-temporal features. We utilize the deformable convolution with residual offsets proposed by~\cite{chanBasicVSRImprovingVideo2022} to warp the features. Inspired by the superior performance of Gated Linear Units (GLUs) in Transformers~\cite{shazeer2020glu} we design and add a GLU based multi-layer-perceptron layer with GeLU activation for efficiently fusing the aligned features extracted from both the noisy and denoised frames.  As shown in \cref{fig:dcgfu_rmdf}(a), we name this deformable convolution-GLU combination as the DC-GFU module. At this fusion-stage each frame is processed separately and the fusion is performed only along the channel dimension. Recurrence is applied for the alignment stage across all the multi-scale features of each frame. For the fusion module, we do not employ recurrence, as different scale features capture distinct long-range dependencies owing to their varying receptive fields.

\textbf{Deep Feature Fusion using Recurrent Multi-scale Residual Dense Feature Fusion Unit}: 
Post warping we want to perform a robust-to-noise dense feature fusion while taking advantage of the temporal correlations among the features of all the input frames and also the spatial correlations between the multi-scale features within the same frame. For this task, we design and propose a Recurrent Multi-scale Residual Dense Feature Fusion Unit (RMDF) as shown in \cref{fig:dcgfu_rmdf}(b). The recurrence comes from the fact that the same RMDF module is applied progressively to all the frames' features. For any frame $t$, the RMDF takes in the corresponding frame's multi-scale features \{$\mathbf{F}^{\,1}_{\,t}$, $\mathbf{F}^{\,2}_{\,t}$, $\mathbf{F}^{\,4}_{\,t}$\}, bi-linearly interpolated noisy frames \{$\mathbf{I}_{\,t}^{\,1}$, $\mathbf{I}_{\,t}^{\,2}$, $\mathbf{I}_{\,t}^{\,4}$\} and a hidden state $\mathbf{h}_{\,t-1}$ as inputs. The multi-scale features are progressively fused in a feed-forward fashion, with Residual Dense Block (RDB) as the basic block, to effectively extract both the short and long range dependencies required for good reconstruction. As shown in \cref{fig:dcgfu_rmdf}(b), multi-scale features are extracted from the noisy frames and fused with the other corresponding input features to minimize any errors accumulated through the previous stages. While these features are utilized to exploit the spatial correlations within the frame, the hidden state $\mathbf{h}$ captures the temporal correlations between all the input frames. Thus, the design of RMDF enables it to extract densely fused multi-scale spatio-temporal features required for enhanced quality outputs. 

\textbf{Multi-Scale Reconstruction using Residual Frame Refinement Module}:
\begin{figure}[tb]
  \centering
   \includegraphics[width=\linewidth]{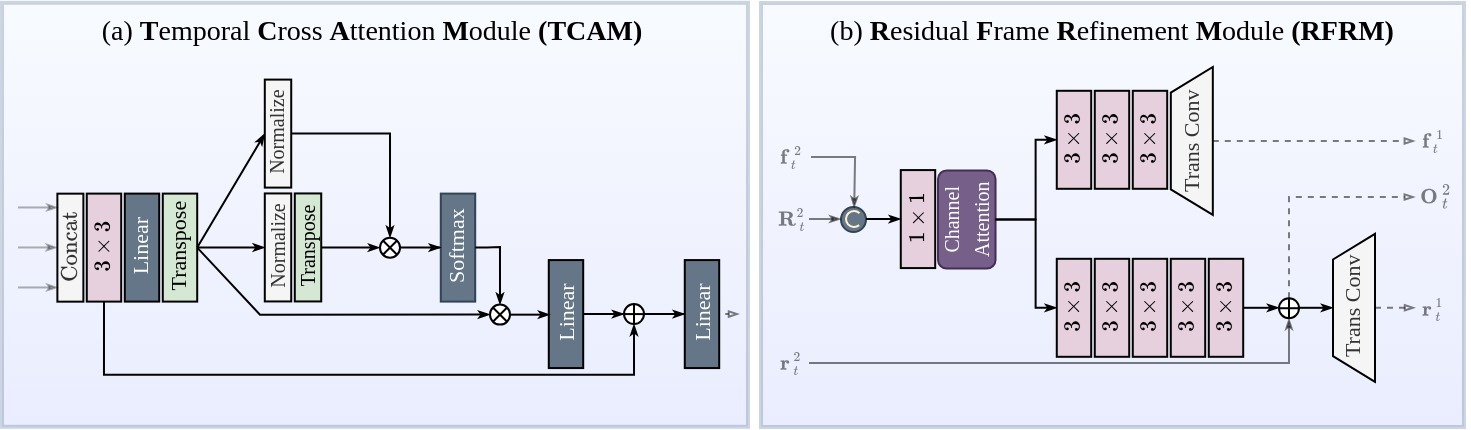}
   \caption{Design of the proposed modules TCA and RFRM.}
   \label{fig:tca_rfrm}
\end{figure}
Considering the heavy noise in the input quanta frames, this ill-posed problem's restored image subspace can be quite large. To output a restored image close to the ground truth we prefer deep supervision that lets the model preserve critical details of the scene. We opt for a multi-scale reconstruction approach where the image at each scale is reconstructed in a progressive fashion. Experiments reveal the efficacy of this approach (\cref{fig:quiver_outline}, \cref{fig:stages_importance}). The stage starts with the lowest scale features extracted from all the frames using RMDF being concatenated and sent into a newly designed Temporal Cross Attention (TCA) module. As shown in \cref{fig:tca_rfrm}(a), the TCA module is similar to the multi-head attention in vision transformers~\cite{dosovitskiy2020image} in terms of generating queries, keys, and values. However, we maintain the number of heads to be one and apply attention only on the channel dimension. The cross attention is from the fact that we input features extracted from all the input frames. The extracted cross-attention features are then fed into a newly designed Residual Frame Refinement Module (RFRM). As shown in \cref{fig:tca_rfrm}(b), RFRM takes in a residual frame $\mathbf{r}^{\,\alpha}_{\,t}$, a hidden state $\mathbf{f}^{\,\alpha}_{\,t}$, and the features as input, concatenates the hidden state with the features to input into the channel attention block~\cite{zhangImageSuperResolutionUsing2018} to emphasize critical spatio-temporal information. Further, we divide the module into 2 branches. While the former is designed to extract multi-scale spatial correlation information and output a modified hidden state $\mathbf{f}^{\,\alpha/2}_{\,t}$, the latter focuses on refining the residual frame to output a corresponding scale reconstructed image $\mathbf{O}^{\,\alpha}_{\,t}$ and a residual frame $\mathbf{r}^{\,\alpha/2}_{\,t}$. The main purpose of this setup is to initially restore the high-level features through estimating $\mathbf{O}^{\,4}_{\,t}$, followed by focusing on the low-level intricate details while refining the residual frames for scales $2$ and $1$. 

\textbf{Loss Function}. We train QUIVER with a multi-scale loss. The overall loss function can be represented as
\begin{align}
   \mathcal{L}_{\text{\tiny{Q}}} = \; \lambda_1\cdot\mathcal{L}(\mathbf{I}^{\,1,\text{GT}}, \mathbf{I}_{\,d}^{\,1}) + \lambda_2\cdot\mathcal{L}(\mathbf{I}^{\,1,\text{GT}}_{\,t}, \mathbf{O}_{\,t}^{\,1}) + \lambda_3\cdot\mathcal{L}(&\mathbf{I}^{\,2,\text{GT}}_{\,t}, \mathbf{O}_{\,t}^{\,2}) \,+ \cdots \notag \\
   &\lambda_4\cdot\mathcal{L}(\mathbf{I}^{\,4,\text{GT}}_{\,t}, \mathbf{O}_{\,t}^{\,4}),
\label{eq:loss_function}
\end{align}
where $\mathbf{I}^{\,\alpha,\text{GT}}_{\,t}$ is the captured $\text{t}^{\text{th}}$ ground truth frame bicubically downsampled by $\alpha$, and $\mathcal{L}(\mathbf{I}_a,\mathbf{I}_b) = ||\mathbf{I}_a - \mathbf{I}_b||_1 + ||\nabla_x\mathbf{I}_a - \nabla_x\mathbf{I}_b||_1 + ||\nabla_y\mathbf{I}_a - \nabla_y\mathbf{I}_b||_1$. Here, $\nabla_x$ and $\nabla_y$ represent the operations of computing horizontal and vertical gradients.

\section{Proposed I2-2000FPS dataset}
\begin{figure}[tb]
  \centering
   \includegraphics[width=0.85\linewidth]{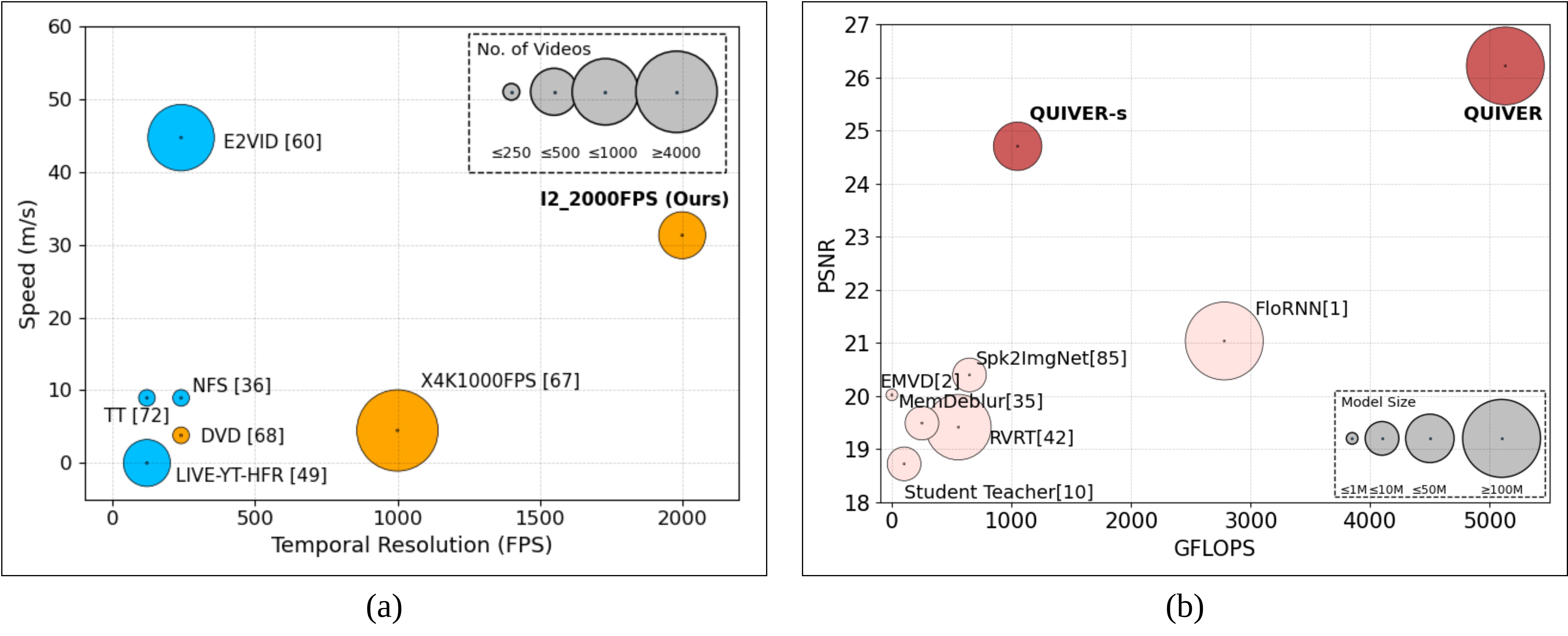}
   \caption{\textbf{(a) Benchmarking high-speed video datasets}. Horizontal axis represents the temporal resolution and the vertical axis indicates the maximum speed captured by the dataset, assuming a fixed camera-object distance. The circles in blue and orange indicate blur and blur-free videos, respectively. \textbf{(b) Benchmarking Quanta Video Restoration Models} on the I$2$-$2000$FPS dataset. Horizontal axis represents the computational complexity in terms of GFLOPs and the vertical axis indicates the PSNR acquired at $3.25$PPP.}
   \label{fig:data_comp}
\end{figure}

While several high-frame-rate datasets have been open-sourced in recent times~\cite{simXVFIEXtremeVideo2021, madhusudanaSubjectiveObjectiveQuality2021, voeikovTTNetRealTimeTemporal2020, rebecqHighSpeedHigh2021, suDeepVideoDeblurring2017, kianigaloogahiNeedSpeedBenchmark2017}, these datasets mainly feature videos tainted by severe motion blur, making them unsuitable in our problem setting. Moreover, features such as high motion speed and sufficient number of videos are also not always guaranteed. A visual representation of existing datasets' comparison is shown in \cref{fig:data_comp}(a). To address the gaps, we introduce the I$2$-$2000$FPS dataset, a high frame rate video collection meticulously designed to capture high-speed motion with precision. 

The I$2$-$2000$FPS dataset has a temporal resolution of $2000$ FPS and a spatial resolution of $512\times1024$ pixels, comprising $280$ unique videos spanning $114$ diverse scenes. The videos are captured using the Chronos $1.4$ high-speed CMOS sensor-based camera from Kron Technologies. 
Notably, I$2$-$2000$FPS incorporates dark current calibration, leveraging the camera's capabilities to mitigate dark current effects. Throughout the data collection process, analog and digital gain were consistently maintained at $0$ dB to avoid amplification of noise. To minimize noise, the videos are exclusively captured outdoor with ambient lighting conditions. \textit{More details on I$2$-$2000$FPS can be found in the supplementary.}

\section{Experiments}
\subsection{Image Formation Model}
\label{subsec:QIS_model}
For experiments involving synthetic data, we use a single-photon detector simulator based on an underlying image formation model discussed below. We build upon the prototype initially suggested in~\cite{maPhotonnumberresolvingMegapixelImage2017} adopted in prior works~\cite{fengyangBitsPhotonsOversampled2012, gnanasambandamHDRImagingQuanta2020, gnanasambandamExposureReferredSignaltoNoiseRatio2022, chanImagesBitsNonIterative2016b, chiDynamicLowLightImaging2020, chiHDRImagingwithSpatiallyVarying2023, quSpatiallyVaryingExposure2024}.
 
Given the quanta exposure~\cite{fossum2013modeling}, $\mathbf{I}^{\text{GT}}$, dependent on the photon flux and exposure time, the observed signal by the sensor can be represented as a Poisson-Gaussian random variable, where the Poisson represents the photon arrival process and the Gaussian models the read noise. The readout process involves various sources of distortions and an Analog-to-Digital Converter (ADC) to convert the real numbers into integers  $\{0,1,2,...,\text{L}\}$, where $L = 2^{\text{Nbits}} - 1$ depending on the bit-depth ($\text{Nbits})$ allocated to the sensor. The final sensor readout, $\mathbf{Y}$, can be represented using the following equation,
\begin{align}
    \mathbf{Y} \sim \; \text{ADC}_{\left[0, \text{L}\right]} \{\text{Poisson}(\text{QE} \times \mathbf{I}^{\text{GT}} + \theta_{\text{dark}}) + \underbrace{\text{Gauss}(0, \sigma^2_{\text{read}}\mathbf{1})}_\text{read noise}\} .\label{eq: mean_ph_arrival}
\end{align}
Akin to previous works~\cite{fengyangBitsPhotonsOversampled2012, gnanasambandamHDRImagingQuanta2020, gnanasambandamExposureReferredSignaltoNoiseRatio2022, chanImagesBitsNonIterative2016b, chiDynamicLowLightImaging2020}, we assume our sensor to be monochromatic as we utilize monochromatic real data in our experiments. For our sensor prototype, we utilize a Quantum Efficiency (QE) of $0.80$. The dark current ($\theta_{\text{dark}}$) and read noise ($\sigma_{\text{read}}$) are set to $1.6\,\text{e}^-$/pix/sec and $0.2\,\text{e}^-$/pix, respectively. 
\subsection{Experimental Settings}
\label{subsec:exp_settings}
\textbf{Training data}. We curate a set of $249$ videos from the I$2$-$2000$FPS collection and employ it as the training dataset for all the deep-learning models in our experiments. Each training sample is fetched on the fly from each clip. A training sample here is defined as a tuple containing the ground-truth/target frames and the $3$-bit quanta frames simulated at $3.25$ photons-per-pixel (PPP) ($\sim1$ lux assuming a $1.1 \mu$m pixel pitch and a $1/2000$ second exposure time) using the image formation model described in \cref{subsec:QIS_model}. \\
\textbf{Testing data}. To effectively analyze the performance of various methods, we carefully sample $31$ videos from I$2$-$2000$FPS containing various motion types, shapes, and speeds. To test the generalizability, we also test the algorithms on X$4$K$1000$FPS~\cite{simXVFIEXtremeVideo2021} test dataset containing $15$ videos from distinct scenes. Lastly, to measure the performance on real-world data, we collect binary frames using a SPAD sensor~\cite{duttonSPADBasedQVGAImage2016} and compare the reconstructed outputs. More details will be discussed in \cref{subsec:results}.\\
\begin{figure}[tb]
  \centering
  \includegraphics[width=0.9\linewidth]{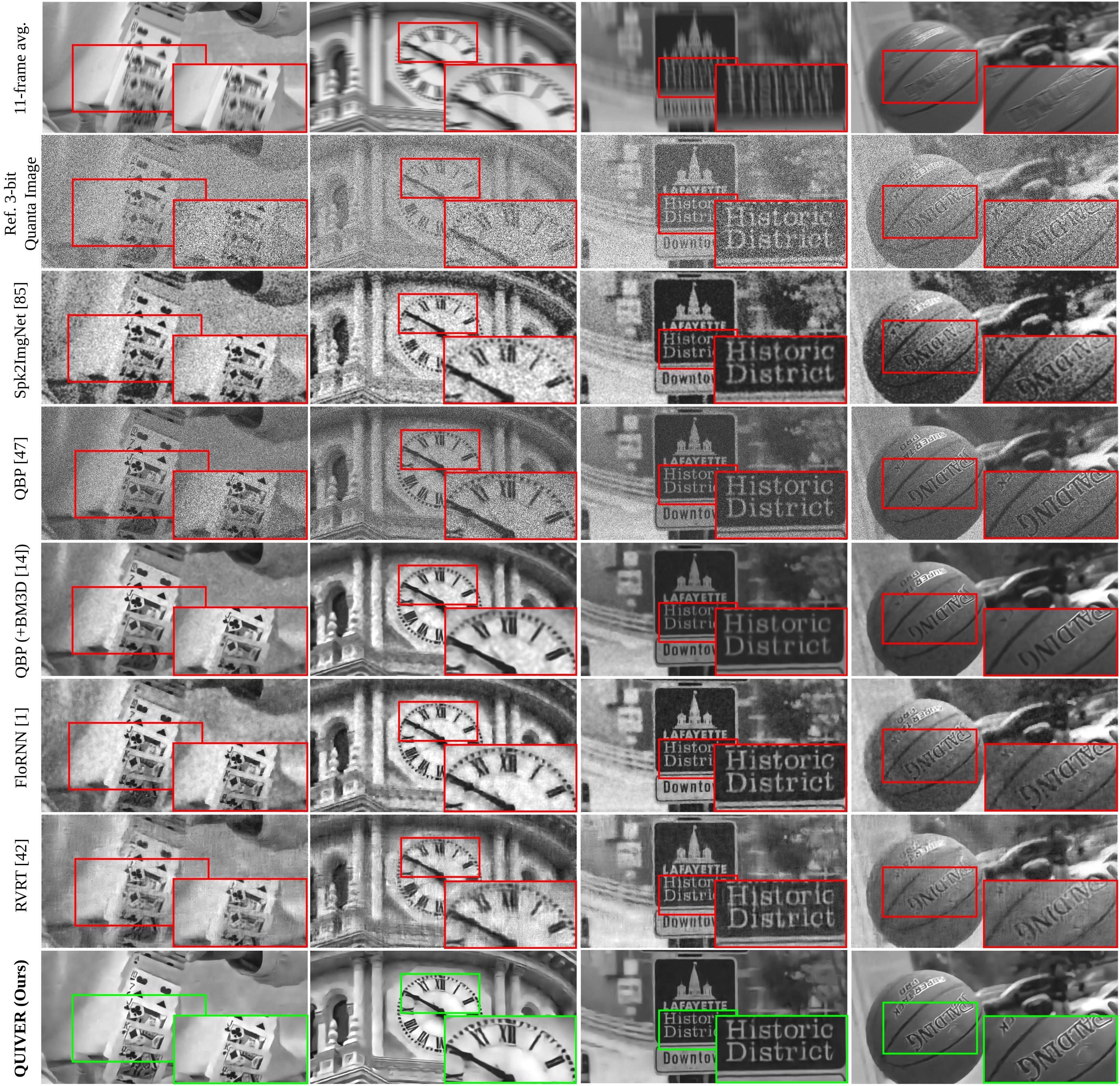}
   \caption{Visual comparisons of the reconstructed results on test videos from the proposed I$2$-$2000$FPS dataset. For fair comparison, all methods utilize $11$ $3$-bit quanta frames simulated at $3.25$ PPP per frame ($\sim 1$ lux) to produce a restored frame. \textit{Best viewed in zoom}.}
   \label{fig:i22000fps_results}
\end{figure}
\textbf{Baselines}. 
We compare the proposed method with eight existing dynamic scene reconstruction algorithms, namely Transform Denoise \cite{chanImagesBitsNonIterative2016b}, QBP \cite{maQuantaBurstPhotography2020}, 
Student-Teacher \cite{chiDynamicLowLightImaging2020}, 
RVRT \cite{liangRecurrentVideoRestoration2022a}, EMVD \cite{maggioniEfficientMultiStageVideo2021}, FloRNN \cite{liUnidirectionalVideoDenoising2022}, MemDeblur \cite{jiMultiScaleMemoryBasedVideo2022}, and Spk2ImgNet \cite{zhao2021spk2imgnet}. We also add an off-the-shelf denoiser BM$3$D \cite{dabovBM3D2007} to QBP, denoted QBP (+BM$3$D), as a baseline for comparison. As we will discuss in \cref{subsec:results}, QUIVER beats all the baselines, both quantitatively and qualitatively. \\
\textbf{Training QUIVER}. We utilize the function mentioned in \cref{eq:loss_function} as the cost function for training QUIVER with regularization parameters $\lambda_1 = 0.2$, $\lambda_2 = 0.85$, $\lambda_3 = 0.1$, and $\lambda_4 = 0.05$. The training data is extracted with patch size $128\times128$ and a batch size of $4$. The weights are initialized with Lecun initialization~\cite{lecunGradientbasedLearningApplied1998}. The network is trained using the Adam optimizer~\cite{kingmadiederikAdamMethodStochastic2015} with an initial learning rate of $2.5\times10^{-5}$. The low learning rate is driven by the inherent instability of recurrent networks, as it mitigates the risk of divergent behavior during training. We use a learning rate scheduler that reduces the learning rate by a factor of $2$ when a plateau is reached. QUIVER takes approximately $1.5$ days to train on a NVIDIA A$100$ Tensor Core GPU using Pytorch.
\begin{figure}[tb]
  \centering
  \includegraphics[width=\linewidth]{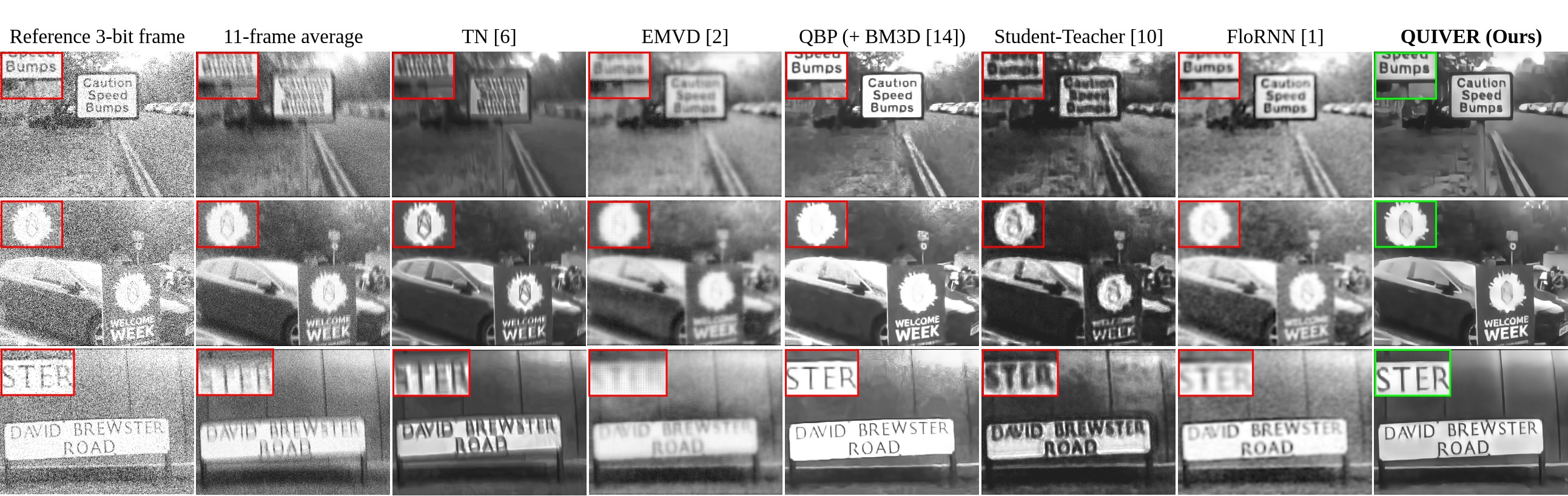}
   \caption{\textbf{Performance on Real Quanta Data}. We capture real $1$-bit quanta data using a SPAD~\cite{duttonSPADBasedQVGAImage2016} and generate $3$-bit frames through temporal averaging. All deep learning based models are trained using a photon level of $4.9$ PPP per frame. \textit{Best viewed in zoom}.}
   \label{fig:real_data_results}
\end{figure}
\subsection{Results}
\label{subsec:results}
\textbf{Synthetic Data Experiments}. We begin with the synthetic experiments where we utilize $3$-bit quanta frames, simulated using the parameters mentioned in \cref{subsec:QIS_model} at $3.25$, $9.75$, $19.5$, and $26$ PPP to test the algorithms' performance. \cref{tab:i22000fps_quant_results} and \cref{tab:x4k1000fps_quant_results} demonstrate the PSNR and SSIM~\cite{wangImageQualityAssessment2004} of various methods extracted by predicting $6017$ I$2$-$2000$FPS frames and $345$ X$4$K$1000$FPS frames. To further substantiate the efficacy of QUIVER’s design, we introduced a scaled-down variant, QUIVER-s (Refer \cref{fig:data_comp}(b) for complexity comparison). Quantitative results indicate that both QUIVER and QUIVER-s offer substantially better performance than all the baselines across a range of light levels. \cref{fig:i22000fps_results} depicts visual results of all the methods on the I$2$-$2000$FPS dataset. It is evident that existing methods fail to handle both motion and noise simultaneously, whereas, our proposed method, QUIVER, produces blur free high SNR outputs while preserving high-frequency details to a large extent. 
\begin{table}[tb]
\caption{Performance comparison on the proposed I$2$-$2000$FPS dataset across various light levels. Models are trained using the I$2$-$2000$FPS dataset. QUIVER performs significantly better than the existing methods.}
    \label{tab:i22000fps_quant_results}
    \centering
    \aboverulesep=0ex
    \belowrulesep=0ex
    \resizebox{0.7\linewidth}{!}{
    \begin{tabular}{c|cccccccc}
    \toprule[2pt]
        Photons-Per-Pixel (PPP) & \multicolumn{2}{c}{$3.25$} & \multicolumn{2}{c}{$9.75$} & \multicolumn{2}{c}{$19.5$} & \multicolumn{2}{c}{$26$}\\ \midrule[0.5pt]
        Method & PSNR$\uparrow$ & SSIM$\uparrow$ & PSNR$\uparrow$ & SSIM$\uparrow$ & PSNR$\uparrow$ & SSIM$\uparrow$ & PSNR$\uparrow$ & SSIM$\uparrow$\\\midrule[1pt]
        Transform Denoise~\cite{chanImagesBitsNonIterative2016b} & $21.3170$ & $0.7184$ & $23.1521$ & $0.7671$ & $22.7748$ & $0.7812$ & $22.3096$ & $0.7811$\\
        QBP~\cite{maQuantaBurstPhotography2020} & $15.9411$ & $0.1293$ & $19.1856$ & $0.2654$ & $20.4000$ & $0.3713$ & $20.7978$ & $0.4114$\\
        QBP (+ BM$3$D~\cite{dabovBM3D2007}) & $21.5476$ & $0.7033$ & $22.2001$ & $0.6899$ & $22.8351$ & $0.7696$ & $22.8617$ & $0.7832$\\
        Student-Teacher~\cite{chiDynamicLowLightImaging2020} & $18.7200$ & $0.4006$ & $16.5195$ & $0.2479$ & $15.7636$ & $0.2133$ & $13.2889$ & $0.0735$\\
        RVRT~\cite{liangRecurrentVideoRestoration2022a} & $19.4115$ & $0.3539$ & $21.6714$ & $0.4568$ & $22.0826$ & $0.5021$ & $21.7528$ & $0.4968$\\
        EMVD~\cite{maggioniEfficientMultiStageVideo2021} & $20.0194$ & $0.5873$ & $21.0559$ & $0.6048$ & $22.4403$ & $0.5592$ & $23.4053$ & $0.5576$\\
        FloRNN~\cite{liUnidirectionalVideoDenoising2022} & $21.0341$ & $0.6785$ & $25.6132$ & $0.7091$ & $\underline{27.4322}$ & $0.7395$ & $27.8520$ & $0.7784$\\
        MemDeblur~\cite{jiMultiScaleMemoryBasedVideo2022} & $19.4877$ & $0.3868$ & $14.4906$ & $0.1112$ & $16.1775$ & $0.1667$ & $16.0058$ & $0.1712$\\
        Spk2ImgNet~\cite{zhao2021spk2imgnet} & $20.3945$ & $0.5642$ & $19.6665$ & $0.6733$ & $22.9372$ & $0.7008$ & $14.9769$ & $0.6861$\\
        \rowcolor{lightgray}
        QUIVER-s (Ours) & $\underline{24.7013}$ & $\underline{0.7565}$ & $\mathbf{26.8676}$ & $\underline{0.7883}$ & $27.2989$ & $\underline{0.8432}$ & $\underline{27.8659}$ & $\underline{0.8408}$\\
        \rowcolor{lightgray}
        QUIVER (Ours) & $ \mathbf{26.2143}$ & $\mathbf{0.7897}$ & $\underline{26.8058}$ & $\mathbf{0.8250}$ & $\mathbf{27.7538}$ & $\mathbf{0.8563}$ & $\mathbf{27.9377}$ & $\mathbf{0.8446}$\\ \bottomrule[2pt]
    \end{tabular}
    }
\end{table}
\begin{table}[tb]
\caption{Performance comparison on the X$4$K$1000$FPS dataset across various light levels. Models are trained using the I$2$-$2000$FPS dataset. QUIVER performs significantly better than the existing methods.}
    \label{tab:x4k1000fps_quant_results}
    \centering
    \aboverulesep=0ex
    \belowrulesep=0ex
    \resizebox{0.7\linewidth}{!}{
    \begin{tabular}{c|cccccccc}
    \toprule[2pt]
        Photons-Per-Pixel (PPP) & \multicolumn{2}{c}{$3.25$} & \multicolumn{2}{c}{$9.75$} & \multicolumn{2}{c}{$19.5$} & \multicolumn{2}{c}{$26$}\\ \midrule[0.5pt]
        Method & PSNR$\uparrow$ & SSIM$\uparrow$ & PSNR$\uparrow$ & SSIM$\uparrow$ & PSNR$\uparrow$ & SSIM$\uparrow$ & PSNR$\uparrow$ & SSIM$\uparrow$\\\midrule[1pt]
        Transform Denoise~\cite{chanImagesBitsNonIterative2016b} & $19.6255$ & $0.6323$ & $22.1703$ & $\underline{0.7044}$ & $22.9938$ & $0.7229$ & $22.6230$ & $0.7204$\\
        QBP~\cite{maQuantaBurstPhotography2020} & $15.5634$ & $0.2302$ & $16.9758$ & $0.3230$ & $17.1798$ & $0.3957$ & $17.7807$ & $0.4188$\\
        QBP (+ BM$3$D~\cite{dabovBM3D2007}) & $17.9677$ & $0.5123$ & $18.5308$ & $0.5226$ & $18.2407$ & $0.5414$ & $18.7917$ & $0.5586$\\
        Student-Teacher~\cite{chiDynamicLowLightImaging2020} & $18.8208$ & $0.3652$ & $16.1548$ & $0.2608$ & $14.9359$ & $0.2571$ & $13.9762$ & $0.1186$\\
        RVRT~\cite{liangRecurrentVideoRestoration2022a} & $19.9203$ & $0.3641$ & $21.0781$ & $0.4472$ & $21.4780$ & $0.4925$ & $20.7899$ & $0.4919$\\
        EMVD~\cite{maggioniEfficientMultiStageVideo2021} & $20.5102$ & $0.4836$ & $21.8152$ & $0.5595$ & $22.9440$ & $0.5936$ & $22.4587$ & $0.5860$\\
        FloRNN~\cite{liUnidirectionalVideoDenoising2022} & $20.8283$ & $0.5778$ & $\mathbf{23.5874}$ & $0.6484$ & $\underline{24.3214}$ & $0.6683$ & $\mathbf{25.2483}$ & $0.7170$\\
        MemDeblur~\cite{jiMultiScaleMemoryBasedVideo2022} & $19.5534$ & $0.3642$ & $14.5595$ & $0.2203$ & $16.6749$ & $0.3116$ & $15.6496$ & $0.2974$\\
        Spk2ImgNet~\cite{zhao2021spk2imgnet} & $18.9424$ & $0.4731$ & $19.2532$ & $0.5722$ & $20.3442$ & $0.5716$ & $16.0931$ & $0.6106$\\
        \rowcolor{lightgray}
        QUIVER-s (Ours) & $\underline{20.9197}$ & $\underline{0.5955}$ & $21.7990$ & ${0.6523}$ & ${24.1924}$ & $\underline{0.7316}$ & ${23.4411}$ & $\underline{0.7248}$\\
         \rowcolor{lightgray}
        QUIVER (Ours) & $\mathbf{21.8730}$ & $\mathbf{0.6521}$ & $\underline{23.1654}$ & $\mathbf{0.7057}$ & $\mathbf{24.5956}$ & $\mathbf{0.7645}$ & $\underline{25.0086}$ & $\mathbf{0.7513}$\\ \bottomrule[2pt]
    \end{tabular}
    }
\end{table}

\textbf{Real Data Experiments}. 
We verify the methods' performance on real data. The real data is collected as binary frames using a SPAD sensor~\cite{duttonSPADBasedQVGAImage2016} at $10000$ FPS with a spatial resolution of $240\times320$. As SPADs possess zero read noise, the binary frames are summed up to generate $3$-bit frames. The average observed light level after summation is $4.9$ PPP. We generate results with networks trained at $4.9$ PPP and demonstrate the visual results in \cref{fig:real_data_results}. QUIVER, as opposed to existing state-of-the-art, effectively recovers high-frequency information while applying visually appealing smoothening effect to low-frequency regions of the scene. It is noteworthy that SPADs' image formation model is significantly different from that of the QIS' imaging model~\cite{rappDeadTimeCompensation2019, rappFewPhotonsMany2017}. Therefore, the visual results also indicate that the proposed QUIVER can thoroughly generalize to various single-photon detectors.
\section{Ablation Study}
\begin{figure}[tb]
    \centering
    \includegraphics[width=\linewidth]{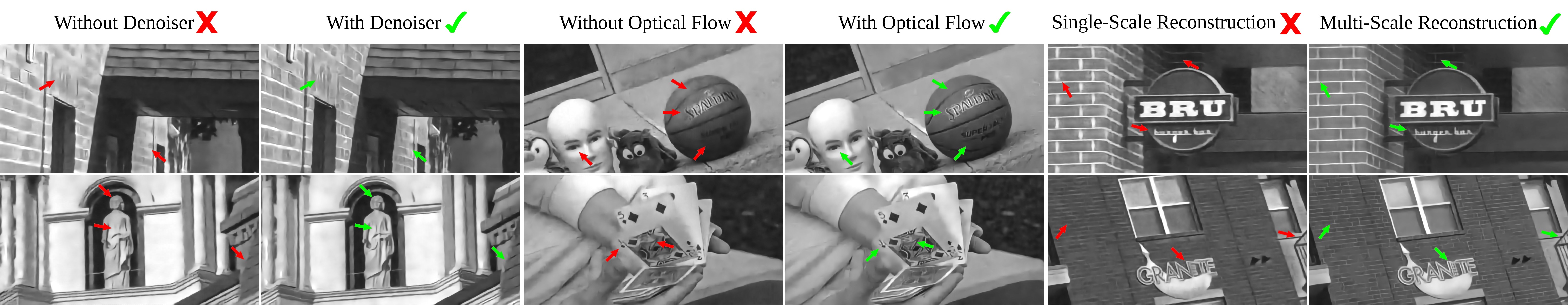}
    \caption{\textbf{Ablation Study}. Visual Comparisons depicting the effectiveness of Pre-denoiser, Optical Flow, and Multi-Scale Reconstruction Modules. \textit{Best viewed in zoom}.}
    \label{fig:stages_importance}
\end{figure}

\textbf{Effect of Pre-Denoising, Optical Flow and Multi-Scale}. We conduct an ablation study to evaluate the effect of the pre-denoising, the learnable optical flow, and multi-scale reconstruction on performance. Upon removal of either module, we expand the features dimension of layers in the feature extraction stage and add RDB blocks to the RMDF module, thereby maintaining a similar model capacity. We train all possible combinations and display the quantitative results in \cref{tab:ablation}. Results indicate that, in the absence of either one or more modules, the network's performance is substantially worse. Visual Results in \cref{fig:stages_importance} indicate that these modules serve critical roles as they significantly contribute to model's performance.

\textbf{Does loading pre-trained optical flow module weights help?} We initialize SPyNet with its pre-trained weights and finetune the same while training QUIVER, and display its quantitative results in \cref{tab:ablation}. As the pre-trained SPyNet is not robust to photon shot noise and read noise, especially in low-light conditions, initializing the module with it will result in sub-optimal performance.
\begin{table}
\caption{\textbf{Ablation study}. We conduct experiments to emphasize the role of Denoiser, Optical flow, and Multi-Scale reconstruction modules. We also show the effect of loading pretrained optical flow weights ($^*$) on performance.}
    \centering
    \resizebox{0.45\linewidth}{!}{
    \begin{tabular}{ccccc}
    \toprule[2pt]
        \multirow{2}{*}{Pre-Denoising} & \multirow{2}{*}{Optical Flow} & \multirow{2}{*}{Multi-Scale} & \multicolumn{2}{c}{I$2$-$2000$FPS}\\ \cmidrule(lr){4-5}
        & & & PSNR$\uparrow$ & SSIM$\uparrow$\\\midrule[1pt]
        \xmark & \xmark & \xmark & $23.6702$ & $0.7756$\\
        \cmark & \cmark$^{*}$ & \cmark & $23.9479$ & $0.7709$\\
        \cmark & \cmark & \xmark & $24.3841$ & $0.7808$\\
        \xmark & \xmark & \cmark & $24.7445$ & $0.7755$\\
        \xmark & \cmark & \cmark & $24.9999$ & $0.7753$\\
        \cmark & \xmark & \cmark & $25.7521$ & $0.7760$ \\
    \rowcolor{lightgray}
        \cmark & \cmark & \cmark & $\mathbf{26.2143}$ & $\textbf{0.7897}$\\
    \bottomrule[2pt]
    \end{tabular}}
    \label{tab:ablation}
\end{table}

\section{Conclusion}
In this paper, we presented a methodology to reconstruct blur-free grayscale images/videos captured using $1$-bit or few-bit quanta data. While adopting the ideology of classical quanta restoration methods, we proposed an end-to-end deep learning framework, QUIVER, that utilizes pre-filtering, learnable optical flow module, and a novel multi-scale reconstruction approach to produce high-visual outputs. Experiments on synthetic and real data indicate QUIVER beats state-of-the-art and can generalize across single-photon sensors. We also introduce the world's first high-speed video dataset, I$2$-$2000$FPS, that captures fast moving scenes at $2000$ fps, covering wide ranges of motion. We believe that I$2$-$2000$FPS will be a valuable asset for researchers in high-speed motion analysis and other computer vision tasks.

\textbf{Acknowledgements.} The work is supported, in part, by Samsung Research America (SRA), DARPA / SRC CogniSense JUMP 2.0 Center, NSF IIS-2133032, and NSF ECCS-2030570. The authors would like to thank Germán Mora Martín for his help in recording the SPAD data.

\bibliographystyle{splncs04}
\bibliography{main}
\end{document}